\documentclass[conference]{IEEEtran}
\IEEEoverridecommandlockouts
\usepackage[dvipsnames]{xcolor}
\usepackage{graphicx}
\usepackage{cite}
\usepackage{amsmath,amssymb,amsfonts}
\usepackage{algorithmic}
\usepackage{graphicx}
\usepackage{comment}
\usepackage{textcomp}
\usepackage{xcolor}
\def\BibTeX{{\rm B\kern-.05em{\sc i\kern-.025em b}\kern-.08em
    T\kern-.1667em\lower.7ex\hbox{E}\kern-.125emX}}
\begin{document}

\title{Collusion Attacks on \\ Decentralized Attributed-Based Encryption: \\ Analyses and a Solution}

\author{
\IEEEauthorblockN{Ehsan Meamari$^1$~~~~ Hao Guo$^1$ ~~~~ Chien-Chung Shen$^1$~~~~~~Junbeom Hur$^2$}
\IEEEauthorblockA{$^1$Department of Computer and Information Sciences, University of Delaware, U.S.A. \\
$^2$Department of Computer Science and Engineering, Korea University, South Korea \\
\{ehsan,haoguo,cshen\}@udel.edu, jbhur@isslab.korea.ac.kr}}

\maketitle

\begin{abstract}
Attribute-based Encryption (ABE) is an information centric security solution that moves beyond traditional restrictions of point-to-point encryption by allowing for flexible, fine-grain policy-based and content-based access control that is cryptographically enforced. As the original ABE systems are managed by a single authority, several efforts have decentralized different ABE schemes to address the key escrow problem, where the authority can issue secret keys to itself to decrypt all the ciphertext. However, decentralized ABE (DABE) schemes raise the issue of collusion attacks. In this paper, we review two existing types of collusion attacks on DABE systems, and introduce a new type of collusion among authorities and data users. We show that six existing DABE systems are vulnerable to the newly introduced collusion and propose a model to secure one of the DABE schemes.
\end{abstract}


\section{Introduction}

For Alice to communicate securely with Bob via public-key encryption, she encrypts a message with Bob's public key, which Bob decrypts with his private key. Here Alice knows that it is Bob who she would like to communicate with. However, there are situations where a data sender (data owner or DO) would like to share data securely with (multiple) data receivers (data users or DU) whose identities are not known at the time of sharing.
Although their identities are not known to the sender ahead of time, the intended receivers or interested users could be characterized by certain {\em attributes}. For instance, in the context of medical research, a patient may want to share his/her medical information with receivers who possess attributes such as [``Doctor" or (``Researcher" \& ``Pathology Department")].
As the patient may not have the identities of all the eligible data users, the conventional two-party public-key encryption schemes cannot be applied.

In 2005, Sahai and Waters \cite{Sahai-2005-Introducing-ABE} introduced the concept of Attribute-Based Encryption, which allows the mentioned patient to share his/her medical information with all the eligible data users without knowing their explicit identities. 
There are two kinds of ABE systems: Ciphertext Policy Attribute-based Encryption (CP-ABE) \cite{Bethencourt-First-CP-ABE} and Key Policy Attribute-based Encryption (KP-ABE) \cite{Goyal-2006-First-KP-ABE}. In CP-ABE, each data user receives one secret key for each of his/her attributes from the Authority. The sender encrypts the data with an {\em access policy} specifying the desired attributes of the intended receivers. Following the previous example, the patient determines that only data users who are either doctors or pathology researchers could decrypt the ciphertext. In contrast, in KP-ABE , a data user receives one secret key which encodes the predefined access policy to decrypt the ciphertext. As CP-ABE is more practical for real world applications, almost all research efforts work on improving CP-ABE. For instance, a more efficient CP-ABE scheme  \cite{Waters-CP-ABE}, privacy-preserving for the data users \cite{6143930,6987293,8116436}, faster decryption \cite{10.1007/978-3-642-40779-6_5}, constant-sized ciphertexts \cite{Wang2013,6406057,Doshi:2014:FSC:2904950.2904978}, revocable data users or attributes \cite{10.1093/comjnl/bxw007, 6883910}, and so on.

The earlier CP-ABE systems are centralized as there is only one authority. The authority decides the system parameters, chooses the master key (MK), and issues Secret Keys (SK) associated with attributes to DU. Such a centralised architecture raises several security issues, such as the single point of failure and the key escrow problem where the authority, by using MK, can generate SK for all the attributes for itself to decrypt ciphertexts. To address these defects, many efforts \cite{Li-Decentralizing-both-Models-of-Water-and-Lewko} have tried to decentralize earlier CP-ABE schemes by ``dividing'' MK among different authorities, requesting a threshold $t$ out of $n$ authorities to cooperate to issue SK, and demanding that none of the authorities can issue SK for itself alone. However, decentralization raises new kinds of collusion attacks.

This paper makes the following contributions. 
We first review two existing types of collusion attacks on DABE, collusion among authorities and collusion among DUs. We then introduce a new kind of collusion where some authorities collude with DUs, so that the colluding authorities can recover the MKs of the other non-colluding authorities. After that, the colluding authorities can take over the entire DABE system and issue new SKs without needing any cooperation or permission from other non-colluding authorities. Furthermore, we show that six of the previously published DABE systems are vulnerable to this new attack. Finally we introduce a model to secure one of the defected systems from the newly introduced of collusion attack.

The paper proceeds in Section~\ref{sec:preliminaries} to review background knowledge. Section~\ref{sec:ColludionAttacks} first discusses two existing collusion models on DABE schemes and then introduces a new collusion model among authorities and DUs. In Section~\ref{sec:VulnerableModels}, we analyze four existing DABE models to show that they are all vulnerable to the newly introduced collusion attack. In Section~\ref{sec:SecuredModel}, we propose a new model to secure one of the vulnerable models. Section~\ref{sec:Conclusion} concludes the paper with future work.

\section{Preliminaries}
\label{sec:preliminaries}

In this section, we review background knowledge needed for the establishment of ABE and DABE systems and the analyses of their security.




\subsection{Bilinear Maps}

Consider $G_0$ and $G_1$ as two multiplicative cyclic groups. Suppose that the prime order of both groups is $p$ and the generator of $G_0$ is $g$. There exists a map $e$: $G_0\times G_0 \rightarrow G_1$, with an efficient algorithm, which, for all $g_1, g_2 \in G_0$, computes $e(g_1, g_2)$. Map $e$ is termed bilinear if it has the following two properties:

1) \textbf{Bilinearity:} For all $g_1, g_2 \in G_0$ and $a, b \in Z_p$, there is equation $e(g_1^a, g_2^b) = e(g_1, g_2)^{ab}$.

2) \textbf{Non-degeneracy:} $e(g, g) \neq 1$.

\subsection{CP-ABE}
A CP-ABE scheme consists of the following four algorithms.

\textbf{Setup:} The authority runs the setup algorithm to select the Public Parameters (PP) of the system, choose the Master Key (MK) for itself, and broadcast the Public Key (PK) to the users.

\textbf{Encryption:} A DO specifies an access policy to determine the needed attributes of DUs for decrypting the ciphertext. The DO then uses PK to encrypt messages specifying the access policy as part of the encryption, and broadcasts the ciphertext.

\textbf{Key Generation:} The authority runs the key generation algorithm which uses MK to issue Secret Keys (SK) to DUs based on the list of attributes of each DU. 

\textbf{Decryption:} A DU uses its SK and PK to decrypt the ciphertext. If a DU possesses enough attributes which are specified in the access policy, the DU can recover the message. If not, the decryption algorithm outputs an error to the DU.

\subsection{Decentralized ABE}

The original ABE models are administrated by a central authority. Such a centralized architecture raises issues such as key escrow  \cite{Bozovic2012-A-Model-based-on-Chase-first-decentralized-(KP-ABE)-when-DO-defines-list-valid-DUs}, key exposure \cite{Ramana2015},  ineligible DUs \cite{Yu:2016:ACA:3090725.3090727}, privacy of DUs \cite{Song-Privacy-Preserving-via-blank-token}, forging signatures \cite{Wang2012}, and scalability \cite{Lewko-Decentralizing-Attribute-Based-Encryption}. To address these issues, several effort \cite{Hur-First-Paper, Hur-Decentralized-ABE-for-Military-Networks, Lewko-Decentralizing-Attribute-Based-Encryption, Li-Decentralizing-both-Models-of-Water-and-Lewko, Guo-Multi-Authority-Attribute-Based-Access-Control-with-SC} have been proposed to decentralize ABE so that the responsibilities of a central authority are divided among multiple authorities.

\section{Collusion Attacks on DABE}
\label{sec:ColludionAttacks}

Although decentralization addresses several issues related to centralized ABE, it raises new issues. In this section, we review two existing collusion attacks on DABE: collusion among different DUs and collusion among authorities, and introduce a new collusion attack model among both authorities and DUs.

\textbf{1) Collusion among DUs:} This type of collusion happens when some data users with different SKs, collude with each other and combine their SKs to decrypt a ciphertext which is not accessible for each of them alone but is accessible for the sum of the SKs \cite{Bethencourt-First-CP-ABE}. For instance, suppose that $DU_A$ has two attributes $\{Att_1, Att_3\}$ and $DU_B$ has one attribute $\{Att_2\}$. Then, a DO encrypts a massage and determines the access policy as $\{Att_1 \& Att_2\}$. None of the two DUs has the needed SKs to decrypt the ciphertext individually. Therefore, they might want to collude each other by sharing their SKs to decrypt the ciphertext with a set of $SKs= \{SK_{Att_1}, SK_{Att_2}, SK_{Att_3}\}$.

Most of the excising ABE and DABE schemes adopt a similar idea to secure systems against collusion among DUs. Authority/authorities should choose different GIDs when running the Key Generating algorithm to issue the SKs. Different SKs with different GIDs could not be combined to decrypt ciphertext. To the best of our knowledge, all the existing ABE and DABE schemes are secure against this kind of collusion attacks.

\textbf{2) Collusion among authorities.} To address the key escrow problem, several efforts \cite{8486938,Nyamsuren2018,Lin-DABE-with-two-Secure-2PC, Sultan2017} had proposed different DABE architectures, in which none of the authorities is able to issue SKs by itself, as long as authorities do not collude each other. However, these efforts had assumed explicitly that authorities behave honestly so that they do not collude with each other nor share their MKs with each other. Since there is no easy way to monitor the authorities for collusion, this is not a acceptable assumption. To the best of our knowledge, there is not a DABE system which could prevent authorities from colluding. Therefore, the key escrow problem remains an open issue.

\textbf{3) Collusion among authorities and DUs}. In a DABE system, each authority has its own MK and is responsible for protecting it from leaking to other authorities. In addition, there should be no chance for other authorities to circumvent the security of an authority to uncover its MK. However, one potential vulnerability is the collusion between other authorities and DUs and the colluding authorities might uncover the MKs of non-colluding authorities. This paper analyzes the limitations of some of the existing DABE schemes to show their vulnerability to this kind of collusion attack.

\section{Analysis of Vulnerable Schemes}
\label{sec:VulnerableModels}

In this section, we analyze four existing DABE models to show that they are vulnerable to the newly introduced collusion attack among authorities and DUs. We show how some of the authorities can collude with one DU to uncover the MK of the other authority. We then describe a solution to secure one of the vulnerable schemes.

Notice that the paper has been using ``multiple authorities'' to represent DABE generically. However, each specific DABE scheme has its unique way of ``decentralizing'' ABE so that the multiple ``authorities'' are not simply replicas, but with different delegated and/or partially replicated functions. In addition, these ``authorities'' are also named differently in different schemes. The following analysis of this paper adopts the specific terminologies used in each scheme. 

\subsection{The Hur Model I}

Hur et al. \cite{Hur-First-Paper} investigated the key escrow problem by decentralizing the Bethencourt model.

\subsubsection{Review of the Hur model I}

Hur et al. developed a DABE model with two authorities: Key Generation Center (KGC) and Attribute Authority (AA). The model works as follows. 

\textbf{Setup:}
First, a trust initializer (TI) decides the public parameters: a bilinear group $G_0$ with prime order $p$ and generator $g$ and a hash function $H: \{0, 1\}^\ast \rightarrow G_0$, and broadcasts public key $PK_{TI} = \{G_0, g, H\}$. Then, KGC chooses a random exponent $\alpha \in_R$\footnote{$x\in_R S$ denotes choosing $x$ randomly from a finite set $S$.} $Z_p^*$, saves its master key $MK_{KGC}= g^\alpha$, and broadcasts its public key $PK_{KGC} = e(g, g)^{\alpha}$.
AA selects a random exponent $\beta \in_R Z_p^*$, saves its master key $MK_{AA} = \beta$, and broadcast its public key  $PK_{AA}=\{g^ \beta, g^ {1/\beta}\}$.


\begin{figure}[h]
\centering
\includegraphics[width=0.48\textwidth]{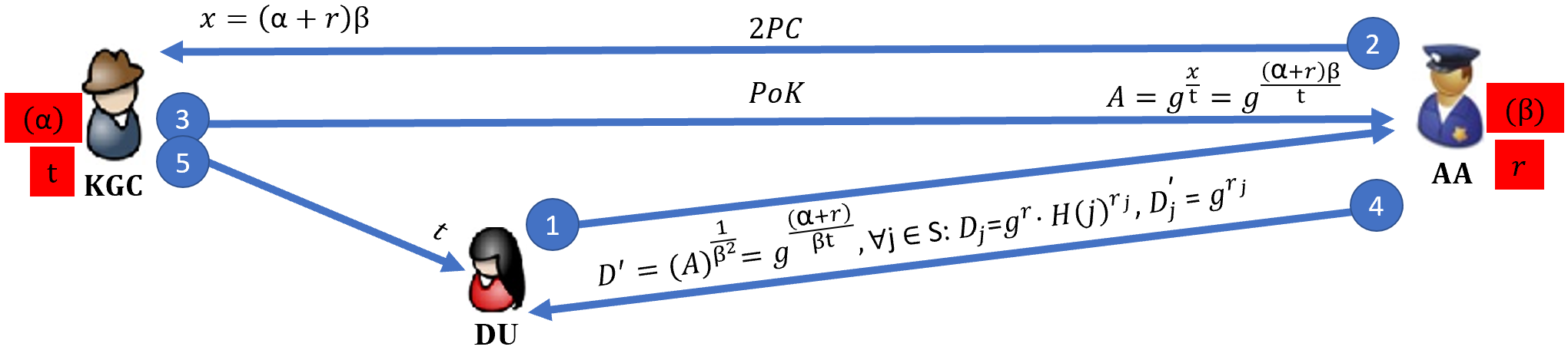}
\caption{Key generation round in the Hur model I}
\label{fig:HurModel1}
\end{figure}

\textbf{Key Generation:}
As depicted in Fig.~\ref{fig:HurModel1}, (1) a DU communicates with the AA to request SKs, based on the attributes which it possesses. (2) AA chooses a specific random exponent $r _p^*$ for the DU which should be unique for each DU to prevent DUs from colluding with each other. Afterward, AA runs a secure two-party computation protocol \cite{10.1007/978-3-642-00468-1_15, Chase-Multi-Authority-(KP-ABE)-Preserves-Privacy-but-Supports-just-AND-policies-Needs-Too-Much-Communications-among-AAs, 10.1007/978-3-540-78524-8_20} to cooperate with KGC to issue the personalized component ($D$) of the SK without leaking their MKs to each other. The secure two-party computation protocol outputs $x=(\alpha+r)\beta$ to KGC. (3) KGC chooses a random exponent $t \in_R Z_p^*$ and sends $A=g^{x/t}$ to AA. (4) KGC shares the choosen $t$ with DU. (5) AA selects random exponent $r_j \in_R Z_p^*$ for each $j \in S$ (the set of DU's attributes), and sends $D'$, a list of $D_j$ and $D'_j$ to  DU. Finally, DU calculates $D=(D')^t$ so that the computed SK would be the same as the one in the Bethencourt model \cite{Bethencourt-First-CP-ABE}.

The Hur model I does not change the formulas used in the Setup and Key Generation algorithms. Instead, the Hur model I divides the responsibilities of the single authority in the Bethencourt model between two authorities KGC and AA. As SK, MK and PK are the same as in \cite{Bethencourt-First-CP-ABE}, the encryption and decryption algorithms remain the same.

\subsubsection{Vulnerability Analysis}

The Hur model I is secure against collusion among different DUs, because AA considers a unique $r$ (as GID) for each DU. However, given that the Hur model I provided no explicit mechanism to prevent collusion between AA and KGC, it is vulnerable to such an attack.

Furthermore, the following analysis demonstrates that the Hur model I is not secure against collusion between DU and AA. Suppose that AA colludes with a DU and receives $t$ from the DU, while it has $A=g^{x/t}$. AA can calculate $({g^{x/t}})^t=g^x$. In addition, AA has $\beta$, and hence can find $1/\beta$ and then $(g^x)^{1/\beta}=g^{\alpha+r}$. Finally, AA divides $g^{\alpha+r}$ by $g^r$ to get $g^\alpha$. AA cannot find $\alpha$ from $g^\alpha$ because solving such a logarithm problem is not easy.

However, AA does not need $\alpha$ as $g^\alpha$ is enough to generate the component parts of SK which are generated by KGC. Therefore, AA can generate SKs for itself or for any other new DU without any permission and cooperation with KGC. For instance, suppose AA decides $r^*$ for a new DU or for itself. Then it calculates $g^\alpha \cdot g^{r^*} = g^{\alpha+r^*}$. It finds $(g^{\alpha+r^*})^{1/\beta}$ which is the $D$ part of the secret key for the new DU that is supposed to be generated through a secure two-party computation protocol with the cooperation of both KGC and AA. Therefore, AA generated  SK without any permission from KGC.


A similar collusion cannot happen between KGC and a DU. Even if KGC colludes with a DU to receive $D'$, $D_j$ and $D'_j$, KGC cannot uncover $\beta$ and $r$ from them. Although KGC cannot launch effective collusion with any DU, it should worry about collusion between AA and a DU. As there is no way for KGC to prevent AA from such a collusion, KGC is obligated to trust AA.


\subsection{The Hur Model II}

Hur et al. \cite{Hur-Decentralized-ABE-for-Military-Networks} extended their earlier work \cite{Hur-First-Paper} to scale the number of authorities.

\subsubsection{Review of the Hur model II}

This model decentralized the Bethencourt model \cite{Bethencourt-First-CP-ABE} by using a central authority (CA) and a set of attribute authorities $A_1, A_2, \cdots, A_m$. This model works as follows.

\textbf{Setup:} First, a trusted initializer chooses a bilinear group $G_0$ of prime order $p$ and generator $g$. In addition, it selects hash functions $H$. Then it broadcasts the public parameter $PP=\{G_0, g, H\}$. Then, CA chooses a random exponent $\beta \in_R Z_R^*$ as its MK and publishes its public key $PK_{CA}=\{g^\beta\}$. Similarly, each $A_i$ selects a random exponent $\alpha_i \in_R Z_R^*$ as its MK and broadcasts its public key $PK_{A_i}=e(g, g)^{\alpha_i}$.

\begin{figure}[h]
\centering
\includegraphics[width=0.48\textwidth]{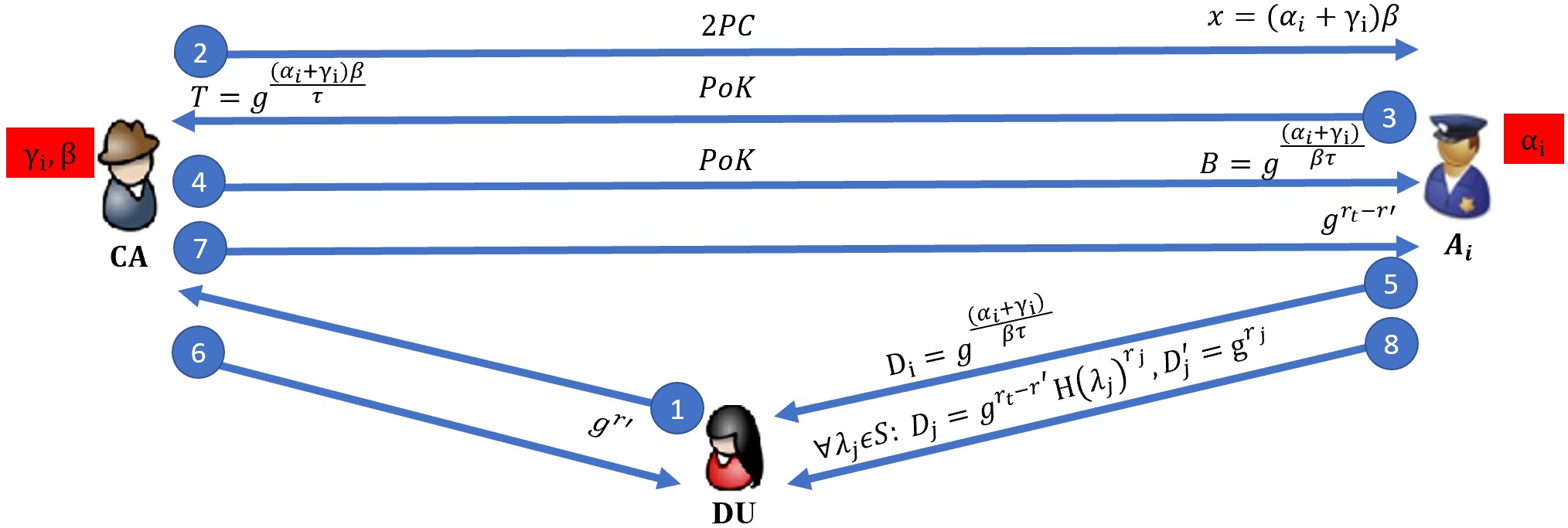}
\caption{Key generation round in the Hur model II}
\label{fig:HurModel2}
\end{figure}

\textbf{Key Generation:} As depicted in Fig. \ref{fig:HurModel2}, (1) DU requests SK from CA. (2) CA chooses random exponents $\gamma_i\in_R Z_p^*$ for each $A_i$ such that $\sum_{i=1}^m {\gamma_i} = r_t$. Then CA runs a secure two-party computation protocol via cooperation with each $A_i$ which outputs $x= (\alpha_i+\gamma_i) \beta$ to $A_i$. (3) $A_i$ randomly chooses exponent $\tau \in_R Z_p^*$, computes $T=g^{x/\tau}$, and sends it to CA. (4) CA computes $B=T^{1/\beta^2}$ and then sends it to $A_i$. (5) $A_i$ computes $D_i=B^{\tau}=g^{(\alpha_i+\gamma_i)/\beta}$ and sends it to DU. The protocol depicted in Fig.~\ref{fig:HurModel2} should be ran between CA and each $A_i$. At the end, DU receives all $D_i$ from all the authorities and computes the $D$ part of its SK via $D=\prod_{i=1}^{m} D_i=g^{(\alpha_1+\cdots+\alpha_m+r_t)/\beta}$. (6) To generate the other parts of SK, CA chooses a random exponent $r' \in_R Z_P^*$ and sends $g^{r'}$ to DU. (7) CA sends $g^{r_t-{r'}}$ to $A_i$. (8) $A_i$ selects $r_j \in_R Z_P^*$ and issues different parts of SK to DU (for each attribute $\lambda_j$ related to the set of attributes of DU which is decided by $A_i$) as follow.

\begin{center}
$\forall \lambda_j \in S: D_j=g^{r_t-{r'}} \cdot H(\lambda_j)^{r_j}, {D'}_j=g^{r_j}$
\end{center}

DU computes $g^{r'} \cdot D_j$ to find its total SK which is the same as the Bethencourt model.

\subsubsection{Vulnerability Analysis} Assume that CA colludes with one DU and receives the $D$ part of SK from DU. Since CA knows $\beta$ and $r_t$, it can compute $g^\alpha=D^\beta/g^{r_t}$ which is enough to issue SK to a new DU without needing to cooperate with any other authorities.

\subsection{The Wang Model}

Wang et al. \cite{Wang2016-Decentralizing-Waters-model-by-2PC} proposed a model which diminishes the key escrow problem based on the Waters model \cite{Waters-CP-ABE}.

\subsubsection{Review of the Wang model}

Two entities termed Key Authority (KA) and Cloud Service Provider (CSP) cooperatively issue SKs for DUs through a secure two-party key generation protocol. The model works as follows.

\textbf{Setup:} The Wang model denotes $G_0$ as a bilinear group of prime order $p$ and generator $g$, and choosew bilinear map $\hat e: G_0\times G_0 \rightarrow G_T$. It also chooses hash function $H:(0,1)^*\rightarrow G_0$ and a set of weights $W=\{w_1, w_2, \cdots, w_n\}$ for the set of attributes $A=\{a_1, a_2, \cdots, a_n\}$.

Afterwards, KA chooses random exponents $\alpha_1, \beta \in Z_p$, saves its master key $MK_{KA}=\{\alpha_1, \beta\}$ and broadcasts its public key $PK_{KA}=\{G_0, g, g^\beta, \hat e(g,g)^{\alpha_1}\}$. Similarly, CSP chooses $\alpha_2 \in Z_p$, saves its master key $MK_{CSP}=\alpha_2$ and broadcasts it public key $PK_{CSP}=\hat e(g,g)^{\alpha_2}$.

\begin{figure}[h]
\centering
\includegraphics[width=0.48\textwidth]{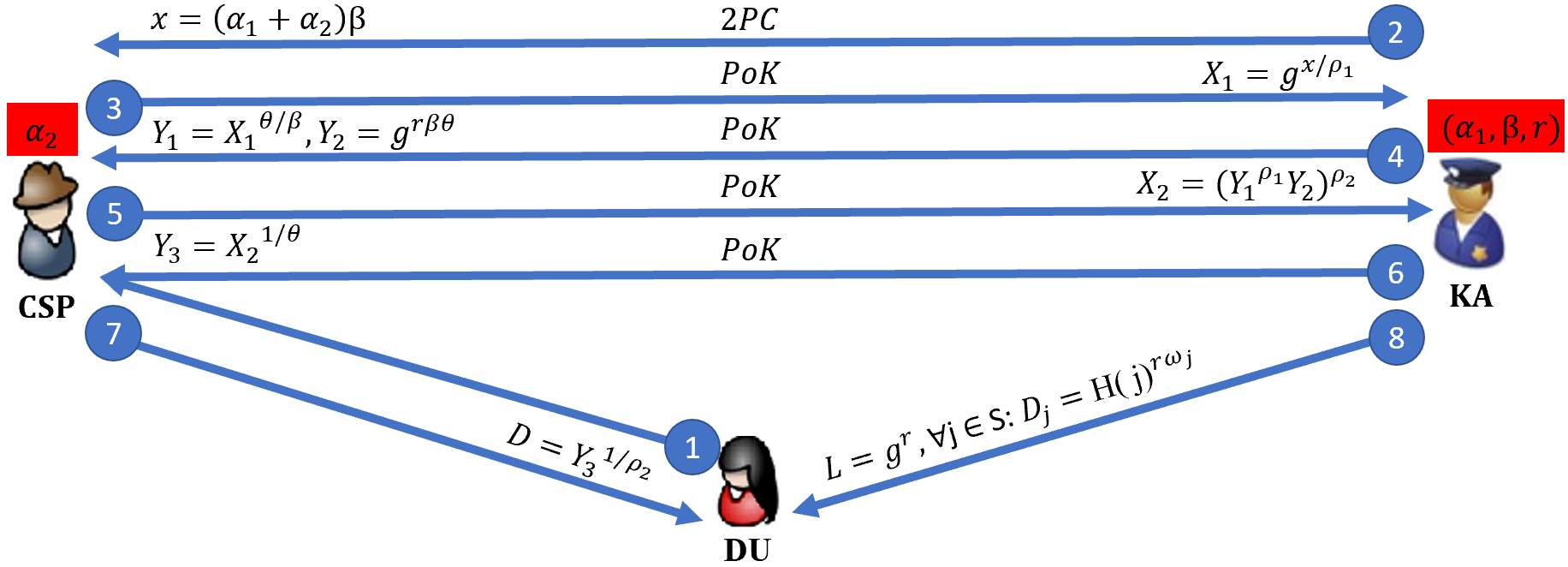}
\caption{Key generation round in the Wang model}
\label{fig:WangModel}
\end{figure}

\textbf{Key Generation:} As depicted in Fig.~\ref{fig:WangModel}, (1) a DU requests SKs for its attributes. (2) KA chooses a unique $r \in Z_p$. Then KA and CSP cooperatively run a secure two-party computation protocol which outputs $x=(\alpha_1+\alpha_2)\beta$ to CSP. (3) CSP chooses a random exponent $\rho_1 \in Z_p$ and sends $X_1=g^{x/{\rho_1}}$ to KA. (4) KA chooses a random exponent $\theta \in Z_p$ and sends $Y_1=X_1^{\theta/\beta}$ and $Y_2=g^{r\beta\theta}$ to CSP. (5) CSP chooses a random exponent $\rho_2 \in Z_p$ and sends $X_2=(Y_1^{\rho_1}Y_2)^{\rho_2}$ to KA. (6) KA computes $Y_3=X_2^{1/\theta}$ and sends it to CSP. (7) CSP issues the $D=Y_3^{1/{\rho_2}}=g^{(\alpha_1+\alpha_2+\beta r)}$ part of SK and sends it to DU. (8) KA issues other parts of SK ($L=g^r$ and $\forall j\in S: D_j={H(j)}^{rw_j}$) and sends them to DU. The complete SK for the DU is as follows.
\begin{center}
$SK=\{D=g^{\alpha_1+\alpha_2+r\beta}, L=g^r, \forall j\in S: D_j={H(j)}^{rw_j} \}$
\end{center}

Since the Wang model is developed based on the Waters model \cite{Waters-CP-ABE}, the general formulas of the setup and key generation algorithms are similar in both models. This similarity concludes similar encryption and decryption algorithms.

\subsubsection{Vulnerability Analysis} At the end of the key generation, DU receives $D=g^{(\alpha_1+\alpha_2+\beta r)}$ from CSP. Now, suppose that KA colludes with one DU and receives $D$ from it. Since KA knows $\alpha_1$ and $\beta$, as its MK, and its chosen $r$, KA computes $D'=g^{(\alpha_1+\beta \cdot r})$ to uncover $g^{\alpha_2} = D/D'$.

Although KA cannot recover $\alpha_2$ from $g^{\alpha_2}$ due to the hardness of the discrete logarithm problem, it doe not need $\alpha_2$. Suppose that a new DU requests SK from KA. Then KA decides $r^*$ as a GID for the new DU and then issues $D=g^{(\alpha_1+\alpha_2+\beta \cdot r^*)}$ without any cooperation with CSP through the secure two-party computation protocol. In addition, KA can issue the other components of SK for all the attributes. Although this model is vulnerable to collusion between KA and DU, there is no chance for CSP to collude with a DU and uncover $\rm{MK_{KA}}$.

\subsection{The Lin Model}

Lin et al. \cite{Lin-DABE-with-two-Secure-2PC} developed a collaborative key management protocol for cloud data sharing.

\subsubsection{Review of the Lin model}

Lin et al. developed a DABE model with three authorities: Key Authority (KA), Cloud Server (CS), and Decryption Server (DS). Both KA and CS issue SKs and DS helps DUs to simplify the decryption process. The model works as follows.

\textbf{Setup:} A TI chooses two multiplicative cyclic groups $G_1$ and $G_2$ with prime order $p$ and generator $g$ of $G_1$. Then it selects two hash functions $H: \{0, 1\}^∗ \rightarrow G_1$ and $H_1: G_2 \rightarrow  Z^∗_p$, chooses a group of random elements $h_1, h_2, \cdots, h_m \in_R G_1$ that are associated with the $m$ attributes, and outputs public parameters $PP = \{g, h_1, h_2, \cdots, h_m, H, H_1\}$. KA chooses a random exponent $q \in_R Z_P^*$ as its master key ($MK_{KA}=q$) and broadcasts its public key $PK_{KA}=g^q$. Similarly, CS chooses a random exponent $\alpha \in_R Z^*_p$, saves its master key ($MK_{CS}=g^\alpha$) and broadcasts its public key $PK_{CS}=e(g, g)^\alpha$.

\begin{figure}[h]
\centering
\includegraphics[width=0.48\textwidth]{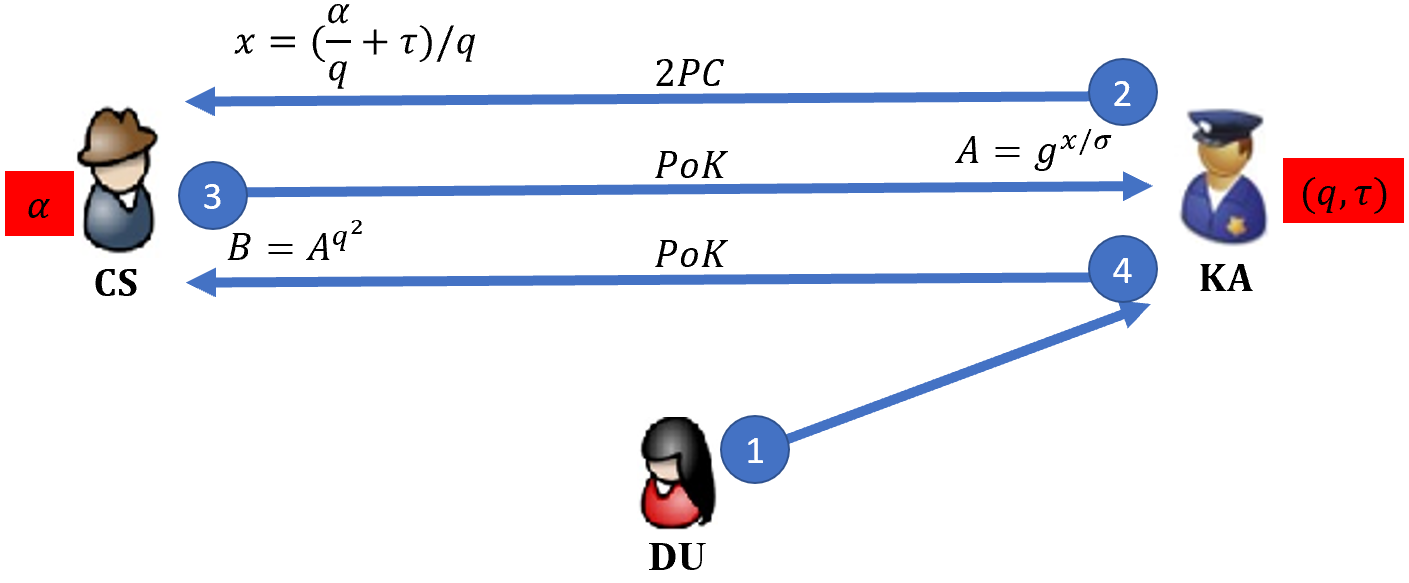}
\caption{The first sub-protocol in key generation round in the Lin model}
\label{fig:LinProtocol1}
\end{figure}

\textbf{Key Generation:} The Key Generation protocol consists of two sub-protocols. In the first sub-protocol, depicted in Fig. \ref{fig:LinProtocol1}, (1) A DU requests SK from  authorities. (2) KA chooses a unique $\tau \in_R Z_P^*$ for DU. Then, CS and KA run a secure two-party computation protocol which outputs $x=(\alpha/q+\tau)/q$ to CS. (3) CS selects a random exponent $\sigma \in_R Z_p^*$ to calculate  $A=g^{x/\sigma}$ and send it to KA. (4) KA calculates $B=A^{q^2}$ and sends it to CS. Then CS calculates and saves $K'=B^\sigma=g^{\alpha+q\tau}$.

\begin{figure}[h]
\centering
\includegraphics[width=0.48\textwidth]{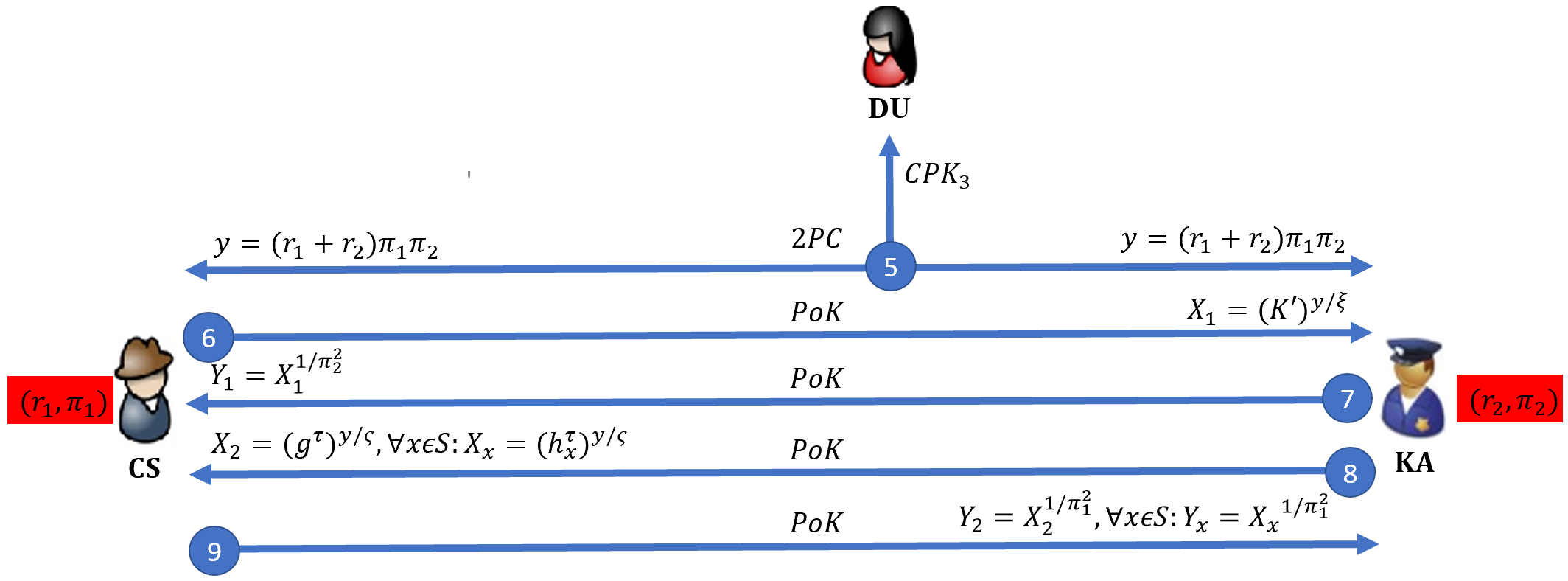}
\caption{The second sub-protocol in key generation round in the Lin model}
\label{fig:LinProtocol2}
\end{figure}

In the second sub-protocol, depicted in Fig. \ref{fig:LinProtocol2},
(5) CS and KA choose random exponents $r_1, \pi_1 \in_R Z^*_p$ and $r_2, \pi_2 \in_R Z^*_p$, respectively. Then, they run another secure two-party computation protocol which outputs $CPK_3=(r1 + r2)/{\pi_1\pi_2}$ to DU, and $y=(r_1 + r_2){\pi_1\pi_2}$ to both CS and KA. (6) CS chooses $\xi \in_R Z_p^*$ and calculates  $X_1=(K')^{y/\xi}$ and sends it to KA. (7) KA computes  $Y_1=X_1^{1/\pi_2^2}$ and sends it to CS. CS computes and saves its secret key $CPK_2= Y_1^{\xi/\pi_1^2}=(g^{\alpha+q\tau})^{(r_1+r_2)/(\pi_1\pi_2)}$. (8) KA selects random exponent $\varsigma \in_R Z_p^*$, calculates  $X_2=(g^\tau)^{y/\varsigma}, \forall x\in S: X_x=(h_x^\tau)^{y/\varsigma}$ and sends it to CS. (9) CS calculates and sends $Y_2=X_2^{1/\pi_1^2}, \forall x\in S:Y_x=X_x^{1/\pi_1^2}$ to KA. Finally, KA calculates and saves its secret key $CPK_1=\{D=Y_2^{\varsigma/\pi_2^2}, \forall x\in S:D_x=Y_x^{\varsigma/\pi_2^2}\}$.

Although the Lin model is developed based on the Water model, the Lin model changes the formulas of SKs as it divides SKs among DU, KA, and CS. Afterwards, DO generates the ciphertext and upload it on CS. Then CS re-encrypts the ciphertext to realize effective attribute revocation.

To decrypt the ciphertext, both KA and CS send their SKs ($CPK_1, CPK_2$) to DS which generates a simpler ciphertext and sends it to DU. DU then uses $CPK_3$ to decrypt the simpler ciphertext.

\subsubsection{Vulnerability Analysis} Suppose that a DU colludes with both KA and DS. Then the DU sends $CPK_3$ to KA and DS sends $CPK_2$, which was received from CS during the decryption round, to KA . Therefore, KA has three secret keys, $CPK_1, CPK_2$, and $CPK_3$. KA can then calculate $K'=CPK_2^{1/CPK_3}=g^{\alpha+q\tau}$. Since, KA has $q$ and $\tau$, it can uncover $g^\alpha=K'/g^{q\tau}$. Although recovering $\alpha$ from $g^\alpha$ is not practical for KA, having $g^\alpha$ is enough to generate SK for a new DU. KA can choose random exponent $\tau^* \in_R Z^*_p$ and generate new $K'=g^{\alpha+q\tau^*}$ without needing to cooperate with CS or running the first sub-protocol in Fig. \ref{fig:LinProtocol1}. Then KA decides random exponents $r^*_1, r^*_2, \pi^*_1, \pi^*_2 \in_R Z^*_p$ and issues $CPK_3=(r^*_1 + r^*_2)/{\pi^*_1\pi^*_2}$ to the new DU, without needing to cooperate with CS or running the second sub-protocol depicted in Fig. \ref{fig:LinProtocol2}. Then, KA issues related $CPK_1$ and $CPK_2$ to itself, while CS does not know anything about newly issued SK.

\subsection{Other Models}

The Hur model I has been adopted as the base model by \cite{Hur-DABE-with-Revocation,Zhao-similar-Hur-first-model-but-discussed-2PC-protocol}, for instance. Although a key revocation capability was introduced in \cite{Hur-DABE-with-Revocation}, due to the same key generation round used as the Hur model I, it suffers from the newly introduced collusion. Similarly, due to the same key generation round used in \cite{Zhao-similar-Hur-first-model-but-discussed-2PC-protocol} as the Hur model I, it is also vulnerable to the newly introduced collusion.

We suspect that other models developed based on the Hur model I/II, the Wang model, and the Lin model might be vulnerable to the collusion attack among authorities and DUs, which need further investigation. 

\section{A Secured Model}
\label{sec:SecuredModel}

In this section, we propose a model to secure the Hur model I against the newly introduced collusion attack.
The secued model decentralizes ABE with two entities, KGC and AA, which works as follow.

\textbf{Setup:} To start the system, a TI chooses the public parameters: a bilinear group $G_0$ with prime order $p$ and generator $g$ and a hash function $H: \{0, 1\}^\ast \rightarrow G_0$, and then broadcasts its public key $PK_{TI}=\{G_0, g, H\}$. Afterwards, KGC chooses two random exponents $\alpha_1, \beta \in_R Z_P^*$, saves its master key $MK_{KGC}=\{ \alpha_1, \beta\}$, and broadcasts its public key $PK_{KGC} = \{g^ \beta, e(g, g)^{\alpha_1}\}$.
Similarly, AA selects a random exponent $\alpha_2 \in Z_P^*$, saves its master key $MK_{AA} =\{\alpha_2\}$, and broadcasts its public key  $PK_{AA}=\{e(g, g)^{\alpha_2}\}$. Since $e(g, g)^{\alpha}=e(g, g)^{\alpha_1}\cdot e(g, g)^{\alpha_2}$ when $\alpha=\alpha_1+\alpha_2$, the public key of the system is as follows.

\begin{center}
$PK=\{G_0, g, H,g^ \beta, e(g, g)^{\alpha}\}$
\end{center}

\begin{figure}[h]
\centering
\includegraphics[width=0.48\textwidth]{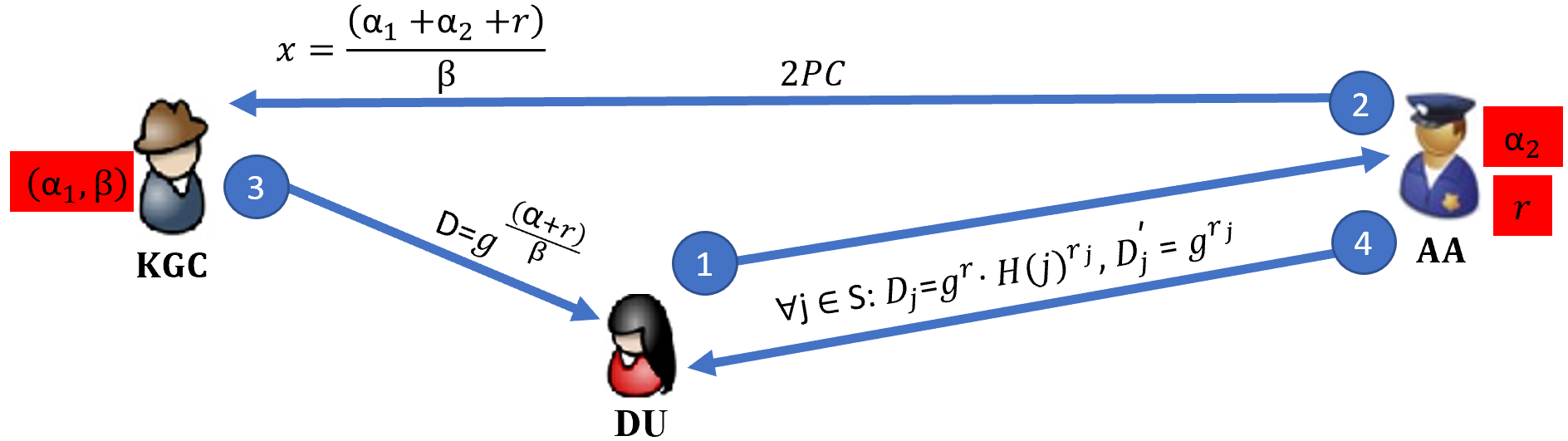}
\caption{Key Generation Round in the Secured Model}
\label{fig:SecuredModel}
\end{figure}

\textbf{Key Generation:} As depicted in Fig.~\ref{fig:SecuredModel}, (1) DU requests SK from AA. (2) AA chooses a specific random exponent $r \in_R Z_P^*$ for the DU. AA then runs a secure two-party computation protocol to cooperate with KGC which outputs $x=1/\beta \cdot (\alpha_1+\alpha_2+r)$ to KGC. (3) KGC computes $D=g^{x}$ and sends it to DU. (4) AA issues other components of the SK based on DU's attributes: $\{\forall j \in S: D_j=g^r \cdot H(j)^{r_j}, D'_j=g^{r_j}\}$. Therefore, the final SK for DU is the same as the Bethencourt model.

{\bf Analysis}: Suppose that a DU colludes with AA and sends $D$ to AA. Since AA does not know the two elements of $\beta$ and $\alpha_1$, it cannot uncover any components of $MK_{KGC}$. Therefore, our proposed model is secure against the newly introduced collusion attack.

\section{Conclusion}
\label{sec:Conclusion}

In this paper, we reviewed two types of existing collusion attacks on DABE schemes, and introduced a new type of collusion attack among authorities and DUs. We then analyzed the vulnerability of four DABE models subject to the newly introduced collusion attack. Based on the analyses, we proposed a new model to secure one of the vulnerable DABE models. Secured solutions to other vulnerable DABE models are left as future work.



\bibliographystyle{IEEEtran}
\bibliography{Reference}

\end{document}